\documentclass[a4paper,11pt]{article}
\pdfoutput=1

\usepackage{jcappub}
\usepackage{aas}
\usepackage{orcidlink}
\usepackage{graphicx}
\usepackage{hyperref}
\usepackage{dcolumn}
\usepackage{multirow}
\usepackage{tabularx}
\usepackage{makecell}
\usepackage{booktabs}
\usepackage{xspace}

\usepackage[nameinlink,noabbrev]{cleveref}
\crefname{equation}{Eq.}{Eqs.}
\crefname{section}{Section}{Sections}
\crefname{figure}{Figure}{Figures}
\crefname{table}{Table}{Tables}
\crefname{appendix}{Appendix}{Appendices}
\Crefname{figure}{Figure}{Figures}
\Crefname{equation}{Equation}{Equations}
\Crefname{section}{Section}{Sections}
\Crefname{table}{Table}{Tables}

\usepackage{amsmath}

\usepackage{bm}
\usepackage{tensor}

\renewcommand{\d}[0]{\mathrm{d}}

\newcommand{\kmsMpc}{\,{\rm km\,s^{-1}\,Mpc^{-1}}}

\newcommand{\uhm}{Department of Physics and Astronomy, University of Hawai`i at M\=anoa, 2505 Correa Rd., Honolulu, HI, 96822}
\newcommand{\asu}{School of Earth and Space Exploration, Arizona State University, Tempe, AZ 85287-6004}

\title{DESI Dark Energy Time Evolution is Recovered by Cosmologically Coupled Black Holes}

\author[1,2]{{Kevin~S.~Croker}\orcidlink{0000-0002-6917-0214},}
\affiliation[1]{\asu}
\affiliation[2]{\uhm}
\emailAdd{kcroker@phys.hawaii.edu}

\author[3]{{Gregory~Tarl\'e}\orcidlink{0000-0003-1704-0781},}
\affiliation[3]{Department of Physics, University of Michigan, 450 Church St., Ann Arbor, MI, 48109}

\author[4]{{Steve~P.~Ahlen}\orcidlink{0000-0001-6098-7247},}
\affiliation[4]{Department of Physics, Boston University, 590 Commonwealth Avenue, Boston, MA 02215}

\author[5]{{Brian~G.~Cartwright}\orcidlink{0009-0003-6667-4729},}
\affiliation[5]{Healthpeak Properties, Inc. 4600 South Sycamore Street, Suite 500, Denver, CO 80237}

\author[2,6]{{Duncan~Farrah}\orcidlink{0000-0003-1748-2010},}
\affiliation[6]{Institute for Astronomy, University of Hawai`i,  2680 Woodlawn Dr., Honolulu, HI, 96822}

\author[7]{{Nicolas Fernandez}\orcidlink{0000-0002-3573-339X},}
\affiliation[7]{NHETC, Department of Physics and Astronomy, Rutgers University, Piscataway, NJ 08854}

\author[1]{{Rogier~A.~Windhorst}\orcidlink{0000-0001-8156-6281}}

\arxivnumber{2405.12282}

\abstract{
  Recent baryon acoustic oscillation (BAO) measurements by the Dark Energy Spectroscopic Instrument (DESI) provide evidence that dark energy (DE) evolves with time, as parameterized by a $w_0 w_a$ equation of state.
  Cosmologically coupled black holes (BHs) provide a DE source that naturally evolves with time, because BH production tracks cosmic star-formation.
  Using DESI BAO measurements and priors informed by Big Bang Nucleosynthesis, we measure the fraction of baryonic density converted into BHs, assuming that all DE is sourced by BH production.
  We find that the best-fit DE density tracks each DESI best-fit $w_0w_a$ model within $1\sigma$, except at redshifts $z \lesssim 0.2$, highlighting limitations of the $w_0w_a$ parameterization.
  Cosmologically coupled BHs produce $H_0 = (69.94 \pm 0.81)~\kmsMpc$, with the same $\chi^2$ as $\Lambda$CDM, and with two fewer parameters than $w_0w_a$.
This value reduces tension with SH0ES to $2.7\sigma$ and is in excellent agreement with recent measurements from the Chicago-Carnegie Hubble Program.
  Because cosmologically coupled BH production depletes the baryon density established by primordial nucleosynthesis, these BHs provide a physical explanation for the ``missing baryon problem'' and the anomalously low sum of neutrino masses preferred by DESI.
The global evolution of DE is an orthogonal probe of cosmological coupling, complementing constraints on BH mass-growth from elliptical galaxies, stellar binaries, globular clusters, the LIGO-Virgo-KAGRA merging population, and X-ray binaries.
A DE density that correlates with cosmic star-formation: 1) is a natural outcome of cosmological coupling in BH populations; 2) eases tension between early and late-time cosmological probes; and 3) produces time-evolution toward a late-time $\Lambda$CDM cosmology different from Cosmic Microwave Background projections.
}

\begin{document}
\maketitle
\flushbottom

\section{Introduction}
It has been nearly a quarter century since two pioneering experiments, the Supernova Cosmology Project \cite{Perlmutter_1999} and the High-Z Supernovae Search Team \cite{Riess_1998}, independently discovered the accelerated expansion of the universe using type Ia supernovae as standard candles.  
The acceleration was attributed to a pervasive form of energy, known as dark energy (DE) \cite{1999ASPC..165..431T}, whose nature has remained elusive.  

In 2006, the Dark Energy Task Force (DETF) \cite{2006APS..APR.G1002A} proposed a comprehensive four-stage experimental program to investigate the nature of DE by determining whether the accelerated expansion is consistent with a cosmological constant, $\Lambda$, whether it evolves, or whether it arises from modifications to General Relativity (GR).  
For dynamic models of DE, the DETF adopted a simple linear parameterization of the equation of state of DE, known as the $w_0 w_a$ parameterization:
\begin{equation}
  w = w_0 + w_a(1-a)\,, \label{eqn:w0wa_model}
\end{equation}
where $a$ is the cosmological scale factor, and $w_0$ and $w_a$ are constants.  
To evaluate the stage of ongoing and proposed experiments, the DETF introduced the reciprocal of the area of the error ellipse enclosing the 95\% confidence limit in the $w_0, w_a$ plane as a Figure of Merit.
The results from Stage III experiments \citep{Rubin2023Union, descollaboration2024dark, Brout_2022} have shown consistency with a cosmological constant but exhibit small ($\lesssim 2.5\sigma$) deviations from $\Lambda$CDM in the $w_0 > -1$, $w_a < 0$ direction.  

Recently, the Dark Energy Spectroscopic Instrument (DESI), the first Stage IV spectroscopic survey, reported significantly enhanced cosmology constraints from Baryon Acoustic Oscillation (BAO) measurements using its first year of data \citep{desicollaboration2024desi}.  
The DESI collaboration finds that adopting a $w_0 w_a$ DE equation of state and combining: DESI-1YR BAO data; Cosmic Microwave Background (CMB) lensing data \citep{PlanckLensing2020}; and Type Ia SNe data from the Pantheon+ \citep{Brout_2022}, Union3\citep{Rubin2023Union} and DES-5YR \citep{descollaboration2024dark} datasets leads to deviations from $\Lambda$CDM by $2.5\sigma$, $3.5\sigma$ and $3.9\sigma$, respectively, indicating a strong preference for $w_0 > -1$ and large $w_a < 0$.
These results can be understood by examining the DE density implied by \cref{eqn:w0wa_model},
\begin{align}
  \rho_\mathrm{DE} \propto \frac{e^{3aw_a}}{a^{3(1 + w_0 + w_a)}}\,.\label{eqn:w0wa_soln}
\end{align}
Note that if $1 + w_0 + w_a > 0$, then the denominator would diverge in the early universe as $a \to 0$.
\begin{figure}
  \centering
  \includegraphics[width=\linewidth]{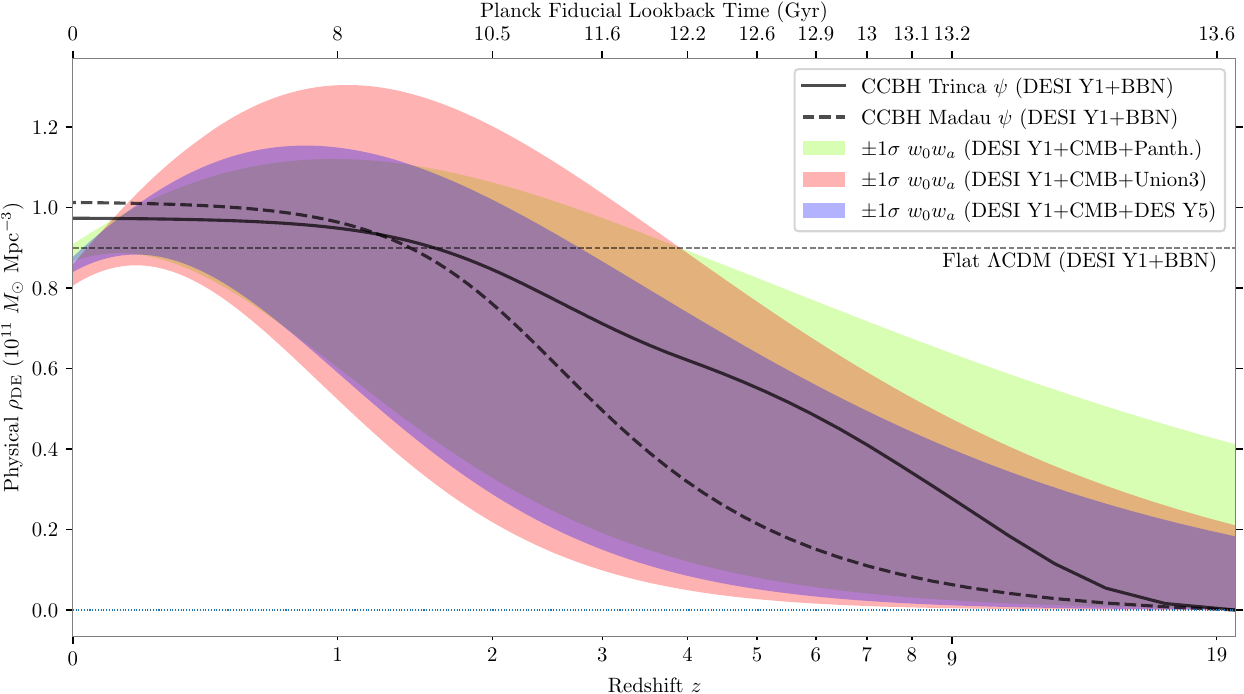}
  \caption{  \label{fig:sfrd}
    Dark energy (DE) density as a function of redshift predicted from cosmologically coupled, stellar-collapse BHs (black).
    Because stellar collapse BH production tracks star-formation, the impact of two empirically determined models is shown.
    The Trinca $\psi$ model (solid) features ample star-formation at high $z$, consistent with recent James Webb Space Telescope observations \cite{Trinca2022, Trinca2024}.
    The Madau $\psi$ model (dashed) is a fiducial rate determined from numerous IR and UV luminosity measurements primarily from $z < 4$ \cite{MadauDickinson2014, madaufragos2017}.
    DESI parameterizes DE time-evolution via equation of state parameters $w_0$ and $w_a$, as determined from joint fits to BAO, CMB, and SNe datasets.
    Regions (shaded) reflect $\pm 1\sigma$ uncertainties in DE density, as determined by SNe datasets from Pantheon (green), Union3 (red), and DES Y5 (blue).
    The time-evolution of DE density, when sourced by stellar-collapse BHs, remains within $<1\sigma$ of the $w_0w_a$ model for each dataset until $z \lesssim 0.2$.
    No combination of $w_0$ and $w_a$ can lead to a DE density that approaches a constant value from below.
    This asymptoting behavior of late-time DE is expected from cosmologically coupled BHs, because the production of stellar progenitors has been decreasing for the past $\sim 9$~Gyr.
  }
\end{figure}%
Thus $1 + w_0 + w_a \leqslant 0$ must always be true, and the DE density initially increases from zero at $a = 0$.
The exponential in the numerator then guarantees a ``hump'' for $w_a < 0$.
Other physical behaviors, like ``asymptotes'' and ``multiple humps,'' cannot be described by \cref{eqn:w0wa_model}.
The DESI data may already highlight the provisional nature of \cref{eqn:w0wa_model}, because the preferred values for $w_0$ and $w_a$ allow causality violation and generate energy densities from nothing.%
\footnote{An equation of state $w < -1$ is often referred to as ``phantom'' for this reason.
Phantom models typically require a negative kinetic energy term, which can be envisioned as a scalar field rolling up its potential \cite{Caldwell:1999ew}. 
In addition, phantom models have been argued to have theoretical inconsistencies \cite{Carroll:2003st, Caldwell:2003vq,Nojiri:2005sx, Fang:2008sn}.  
Because this would imply an unstable vacuum, the interpretation of DESI data within the context of the $\Lambda$CDM model \cite{Colgain:2024xqj, Cortes:2024lgw,Carloni:2024zpl, Wang:2024rjd, Calderon:2024uwn, Park:2024jns}, as well as suggestions for interacting DE \cite{Giare:2024smz}, quintessence scalar fields \cite{Berghaus:2024kra,Tada:2024znt, Yang:2024kdo}, and other scenarios beyond $\Lambda$CDM \cite{Wang:2024hks, Yin:2024hba}, has attracted recent interest from the scientific community.}

The community has established that both of these theoretical obstacles can be removed if DE actually does receive energy from another species.
Models have been proposed where dark matter (DM) couples to DE \cite{Amendola2000, 2004ApJ...604....1F, PereiraJesus2009, CaiSu2010, PourtsidouSkordis2013, Yang:2018euj,Nunes:2022bhn}.  
Very little is known about DM, however, and it has yet to be detected directly \cite{2018PhRvD..98j2006A, 2021PhRvL.126u1803C, 2023PhRvL.131d1003A, 2023PhRvL.131d1002A, 2023PhRvD.107k2013A}, which currently limits the predictive power of these models for the phenomenology of DE.
Another possibility involves models where baryons couple to DE.
For example, the conversion of DE into baryons explains the origin of the Big Bang in the inflationary paradigm \cite{bassett06,2010ARNPS..60...27A}.
The reverse process, the conversion of baryons into DE, was proposed by Gliner to occur during the gravitational collapse of dead stars in the paper that theoretically anticipated DE's discovery \cite{Gliner66}.
Following this study, a diverse community of researchers has developed exact GR solutions of BHs with DE interiors \cite{BlauGuendelman1987, Dymnikova92,  VisserWiltshire04, Lobo06, mazur2015surface} and explored condensed-matter and quantum field-theoretic interpretations (see e.g.~\cite{Chapline01, 2004PNAS..101.9545M, MazurMottola23}).
These BH solutions often mimic the familiar Schwarzschild and Kerr BH models for times $\ll$ the Hubble time, but are singularity-free and need not have horizons.
This can lead to surprising phenomena when these objects are generalized to astrophysically realistic boundary conditions.
For instance, there are known GR solutions for BHs \citep[e.g.][]{FaraoniJacques07, Cadoni23a} that expand in lockstep with their embedding cosmology, gaining mass independently of accretion or merger.%
\footnote{Recent theoretical objections to cosmological coupling \cite{WangWang23, GuarVisser23, Avelino23, Dahal23a, Dahal23b} demonstrate the existence of decoupled solutions.
Because GR is a non-linear theory, the existence of a solution does not imply uniqueness, (see e.g. \S1.2 of \cite{BenderOrszag78}) and so cannot exclude other solutions.
Solutions with realistic cosmological boundary conditions and mass growth $\propto a$ establish these phenomena as a robust GR prediction \cite{Cadoni23a, Cadoni23b}.
Coupling also follows from a mathematically rigorous treatment of the Einstein-Hilbert action for perturbed Robertson-Walker spacetime, where metric degrees of freedom are necessarily constrained in Fourier space \cite{CrokerWeiner19, CrokerRunburg20, CrokerWeiner22}.
It has also been shown that the boundary of a BH must couple to any time-dependent exterior geometry, or else result in a naked null singularity \cite{FaraoniMassimiliano2024}.
This result is independent of the particular field equations, e.g. Einstein's, and applies to any (single) metric theory.}

Recently, observational evidence for cosmological mass growth has been reported in the super-massive black holes (SMBHs) of massive early-type galaxies \cite{Farrah23a, Farrah23b}.
The preferred rate of mass growth for this population, $m \propto a^3$ \cite{Farrah23b} is consistent with the local SMBH mass density proposed by NANOgrav \cite{Lacy23} and low-mass X-ray binaries (XRBs) \cite{GaoLi23}.
Because cosmological number densities decrease $\propto 1/a^3$, if all BHs gain mass $\propto a^3$, then (in the absence of production) \emph{their physical density is constant.}
Conservation of stress energy then requires that they enter Friedmann's equations as DE (i.e. with pressure $P = -\rho$), consistent with the DE-interior BH models.
However, BH growth $\propto a^3$ exceeds upper bounds $\propto a$, determined from known coupled GR solutions and recent measurements in stellar-mass BH populations of LIGO-Virgo-KAGRA \cite{CrokerNishimura20, CrokerZevin21, Ghodla23, Amendola24}, globular clusters \cite{Rodriguez23}, \emph{Gaia} binaries \cite{AndraeElBadry23}, and high-mass XRBs \cite{MlinarZwitter24}.
Local measurements of cosmological mass growth are complicated by uncertainties in accretion, merger history, and formation time, making orthogonal probes of this GR effect valuable for understanding the emerging theoretical and observational landscape. 

DESI and forthcoming Stage IV surveys \cite{LSSTDarkEnergyScience:2018jkl, LSSTDarkEnergyScience:2018yem, LSST:2008ijt, Euclid:2019clj, Amendola:2016saw, Akeson:2019biv, Dore:2019pld,SKA:2018ckk} can provide positive evidence for DE within cosmologically coupled BHs (CCBHs).
The cosmological background offers an averaged dynamics, free from selection effects and assumptions about minimum BH mass%
\footnote{For BH models without horizons, the Tolman-Oppenheimer-Volkoff stability limit $\lesssim 2.2M_\odot$ for maximal neutron star mass can no longer be used to establish a lower-bound BH mass.
This also confounds interpretation of low-mass remnants without  optical or merger tidal-deformation signatures, e.g. \cite{ZevinLed_GW230529}.} 
required to constrain coupled mass growth in isolated stellar binaries.
If BHs contribute as a DE species, then DE time-evolution will track BH production and growth.%
\footnote{The possible role of stellar-collapse BHs in cosmic expansion was first proposed in the Gravitational Aether Theory, an extension to GR \cite{2008arXiv0807.2639A, Prescod-Weinstein2009}.}
In particular, star-formation peaks at $z \sim 2$ \cite{MadauDickinson2014, madaufragos2017} and then drops by a factor of $\sim 10\times$ by today.
This causes the DE density from CCBHs to asymptote toward a constant value at $z \lesssim 1$.
This would mimic $\Lambda$CDM in the late-time universe, \emph{but with a DE density and expansion history that disagrees with CMB projections.}
Because stellar-collapse is expected to be the dominant BH formation channel over the DESI-sensitive redshift range $z \lesssim 3$, and this range overlaps with the peak rate of cosmic star-formation, DESI is well-positioned to distinguish a CCBH population from other hypotheses.

In this paper, we present the first search for significant DE production at ``cosmic noon.''
In \cref{sec:theory} we implement baryon depletion, which tracks the cosmic star-formation rate within Friedmann cosmology.
In \cref{sec:methods}, we detail our methodology for likelihood and parameter estimation with the DESI BAO measurements.
In \cref{sec:results}, we present our main results.
In \cref{sec:discussion}, we discuss implications of cosmologically coupled DE for the current census of baryons in the local universe, the anomalously low summed neutrino mass preferred by DESI, and currently understood BH and accretion physics.
Because we will be describing and comparing different model fit to different datasets, we will consistently refer to these combinations with ``\texttt{Model (Datasets)}'' notation.

\section{Theory}
\label{sec:theory}
We adopt the position that BHs are non-singular vacuum energy objects, cosmologically coupled, and produced solely by stellar collapse \cite[c.f.][]{Privya2024}.
Because the DE density is determined dynamically, defining the cosmology with the density parameters $\Omega_i$ and a distinct present-day Hubble rate $H_0$ is no longer straightforward.
To keep as close to standard notation and usage as possible, we instead focus on ``little $\omega$'s,'' which are typically defined through $\omega_b := \Omega_b h^2$.
This definition implies that,
\begin{align}
  \rho_i\bigg|_{a=1} = \rho_\mathrm{cr} \Omega_i = \left(\frac{3H_0^2}{8\pi G}\right)\Omega_i = \left(\frac{3\times10^4}{8\pi G}\right) \omega_i := C\omega_i.
\end{align}
For matter species, the $C\omega_i$ are comoving energy densities. 
In contrast to $\Omega_i$ and $H_0$, they are directly measured by early-universe probes: $\omega_c$ and $\omega_b$ are the two background parameters of $\Lambda$CDM.
In what follows, we will define $C\omega_b^\mathrm{proj}$ to be the \emph{projected} comoving density of baryons, as inferred from early-universe measurements.
We may now define baryon consumption phenomenologically via the physical baryon density:
\begin{align}
  \rho_b := \begin{cases}
    \displaystyle \frac{C\omega_b^\mathrm{proj}}{a^3} & a < a_i \\
    \displaystyle \frac{C\omega_b^\mathrm{proj}}{a^3} - \frac{\Xi}{a^3}\int_{a_i}^a \psi\frac{\d a'}{Ha'} & a \geqslant a_i
  \end{cases}\,.\label{eqn:rho_b_defn}
\end{align}
Here, $\psi$ is the observed comoving star-formation rate density (SFRD), with units comoving density per unit proper time, consistent with $C\omega_b^\mathrm{proj}$.
Instead, the physical baryon density $\rho_b$ evolves $\propto \omega_b^\mathrm{proj}/a^3$, as inferred by \emph{Planck}, until the onset of stellar first light at $a_i$.
Hereafter, we define $\Xi$ as the fraction of baryons depleted in synchrony with the assembly of stellar mass.
The assumption that $\Xi$ is constant reflects the simplification that its time evolution is subdominant to the time evolution of $\psi$.

We generically define the DE contribution through its pressure as:
\begin{align}
  P_\mathrm{DE} := w \rho_\mathrm{DE} \,, \label{eqn:P_Lambda_defn}
\end{align}
where $\rho_\mathrm{DE}$ is the DE energy density and $w$ its equation of state.
Combining the covariant conservation of stress energy relation $\nabla_\mu \tensor{T}{^\mu_\nu} = 0$, with \cref{eqn:rho_b_defn,eqn:P_Lambda_defn}, we obtain
\begin{equation}
    \frac{\d \rho_\mathrm{DE}}{\d a} + \frac{3}{a}\rho_\mathrm{DE}(1+w) = \frac{\Xi}{Ha^4}\psi \,. \label{eqn:background_cons} 
\end{equation}
The BH DE model is characterized by $w := -1$ and $\Xi \neq 0$.
In this setting, \cref{eqn:background_cons} becomes
\begin{equation}
  \frac{\d \rho_\mathrm{DE}}{\d a} = \frac{\Xi}{Ha^4}\psi\,. \label{eqn:ccbh_de_evolution}
\end{equation}
The Hubble expansion rate $H$ is determined algebraically in terms of the physical densities $\rho_i$ via the Friedmann energy equation,
\begin{align}
  H^2 = \left(\frac{8\pi G}{3}\right)\left[\rho_b(a) + \rho_\gamma(a) + \rho_\nu(a) + \rho_\mathrm{DE}(a) + \rho_c(a)\right], \label{eqn:friedmann}
\end{align}
where $\gamma$, $\nu$ and $c$ denote typical photon, neutrino, and cold dark matter contributions respectively \cite[e.g.][]{dodelsoncosmology2020}.
We omit a curvature contribution $\propto 1/a^2$, so the universe is always spatially flat and critical.
Combined with the derivative of \cref{eqn:rho_b_defn} with respect to $a$, \cref{eqn:ccbh_de_evolution,eqn:friedmann} give a closed system of differential equations for $a \geqslant a_i$
\begin{align}
  \begin{split}
    \frac{\d \rho_\mathrm{DE}}{\d a} &= \frac{\Xi}{Ha^4}\psi\, \\
    \frac{\d \rho_b}{\d a} &= -\left(\frac{3\rho_b}{a} + \frac{\d \rho_\mathrm{DE}}{\d a}\right) \\
    \frac{\d D_\mathrm{M}}{\d a} &= \frac{1}{H a^2}.
  \end{split}. \label{eqn:ode_system}
\end{align}
We include the comoving distance $D_\mathrm{M}$ \cite[e.g.][]{hoggdistance}, as it is required to interpret DESI BAO.
It can be determined by adjusting the integration limits (and sign) using the Fundamental Theorem of Calculus afterward so that $D_\mathrm{M}(1) := 0$.
Note that for flat Robertson-Walker cosmologies, the transverse comoving distance is equal to the line-of-sight comoving distance.
Initial conditions for all species are determined by projecting species' $\omega_j^\mathrm{proj}$ values to $a_i$.
We define the DE density $\rho_\mathrm{DE}(a \leqslant a_i) := 0$.
For $\rho_\nu$, we follow DESI \cite{desicollaboration2024desi} and adopt $N_\mathrm{eff} := 3.044$, a single massive neutrino species with $m_\nu := 0.06~\mathrm{eV}$, and track their energy density exactly \cite[e.g.][\S3.3]{Komatsu2011WMAP}.

\section{Methods}
\label{sec:methods}
\begin{table}
  \centering
  \def\arraystretch{1.2}
  \small
 \begin{tabular}{lcccc}
   \toprule
   \midrule
   Model / Parameter &  \texttt{dynesty} Prior & \texttt{dynesty} Best-Fit & \texttt{dynesty} Posterior & Deviation \\ 
   \midrule
       {\bf CCBH Trinca} $\boldsymbol{\psi}$ && $\left(\chi^2 = 12.66\right)$ & & \\
       $\Xi$ & $\mathcal{U}[0, 10]$ & $1.403$ & $1.396^{+0.050}_{-0.048}$ & $0.15\sigma$ \\
       $\omega_c$ & $\mathcal{U}[0.01, 0.4]$ & $0.1237$ & $0.1240^{+0.0083}_{-0.0078}$ & $-0.04\sigma$ \\
       $100\omega_b^\mathrm{proj}$ & $\mathcal{N}(2.218, 0.055)$ & $2.238$ & $2.219 \pm 0.054$ & $0.35\sigma$\\
       \midrule
           {\bf Flat} $\boldsymbol{\Lambda}${\bf CDM} && $\left(\chi^2 = 12.74\right)$ && \\
           $H_0~[\kmsMpc]$ & $\mathcal{U}[20, 100]$ & $67.72$ & $68.71^{+0.81}_{-0.80}$ & $-1.23\sigma$ \\
           $\omega_c$ & $\mathcal{U}[0.01, 0.4]$ & $0.1129$ & $0.1163^{+0.0082}_{-0.0081}$ & $-0.42\sigma$ \\
           $100\omega_b^\mathrm{proj}$ &$\mathcal{N}(2.218, 0.055)$& $2.101$ & $2.219 \pm 0.055$ & $-2.16\sigma$ \\
           \midrule
   \bottomrule
   \end{tabular}
 \caption{\label{tbl:fit_parameters}
   Fit configuration and results for primary \texttt{dynesty} analysis of DESI Year 1 BAO data.
   Best-fit parameter values, with $\chi^2$, and posteriors distributions at $68\%$ confidence are given.
   Priors are either uniform $\mathcal{U}$ on the indicated range or normally distributed $\mathcal{N}(\mu, \sigma)$.
   Projected comoving baryon density $\omega_b^\mathrm{proj}$ has been multiplied by $100$ for clarity, and is taken from Big Bang Nucleosynthesis (BBN) constraints.
   For the CCBH model, the present-day comoving baryon density is less than $\omega_b^\mathrm{proj}$ due to baryon conversion into DE inside BHs.
   Deviation gives the discrepancy in best-fit values relative to posterior maxima, in units of $\sigma$.
   Note that $\Lambda$CDM best-fit $\omega_b^\mathrm{proj}$ pulls away from the posterior mean, which is dominated by the BBN prior.
   This indicates a preference in the BAO data for lower $\omega_b^\mathrm{proj}$.}
  \vspace{0.1em}
\end{table}%
\begin{figure}
  \centering
  \includegraphics[width=\linewidth]{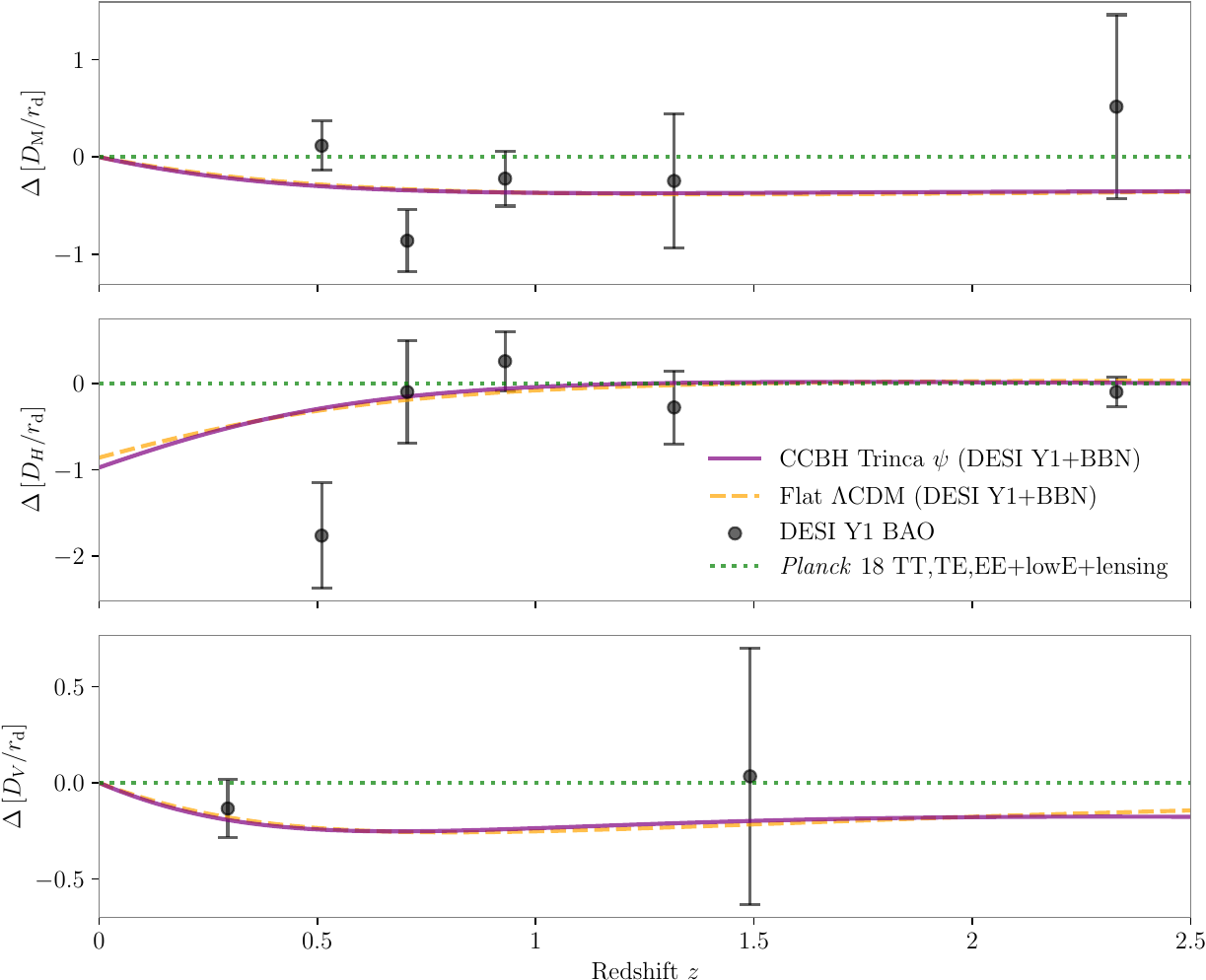}
  \caption{ \label{fig:datafits} 
    Deviations of best-fit models, relative to \emph{Planck} $\Lambda$CDM, in comoving distance $D_\mathrm{M}$ (top), Hubble distance $D_H$ (middle), and a volume-averaged combination $D_V$ (bottom).
    DESI BAO datapoints, relative to \emph{Planck} $\Lambda$CDM, are displayed with $68\%$ confidence (black dots).
    When CCBHs provide the physical source of DE (solid, purple), the resulting expansion history closely tracks a $\Lambda$CDM model (dashed, orange) for $z < 2.5$, but this model is not equal to the $\Lambda$CDM model inferred from early-universe experiments like \emph{Planck} (dotted, green).
    DESI reports that these two $\Lambda$CDM models are discrepant at $1.9\sigma$, qualitatively consistent with a naive $\chi^2 \sim 21$ between the \emph{Planck} best-fit model curve and the 12 DESI BAO degrees of freedom.
    Increased statistics from DESI Y3, Y5, and Y7 datasets, combined with additional constraints from Redshift Space Distortions (not present in the DESI Y1 initial release), should significantly tighten uncertainties and may adjust central values.
  }
\end{figure}%
\begin{figure}
  \centering
  \includegraphics[width=\linewidth]{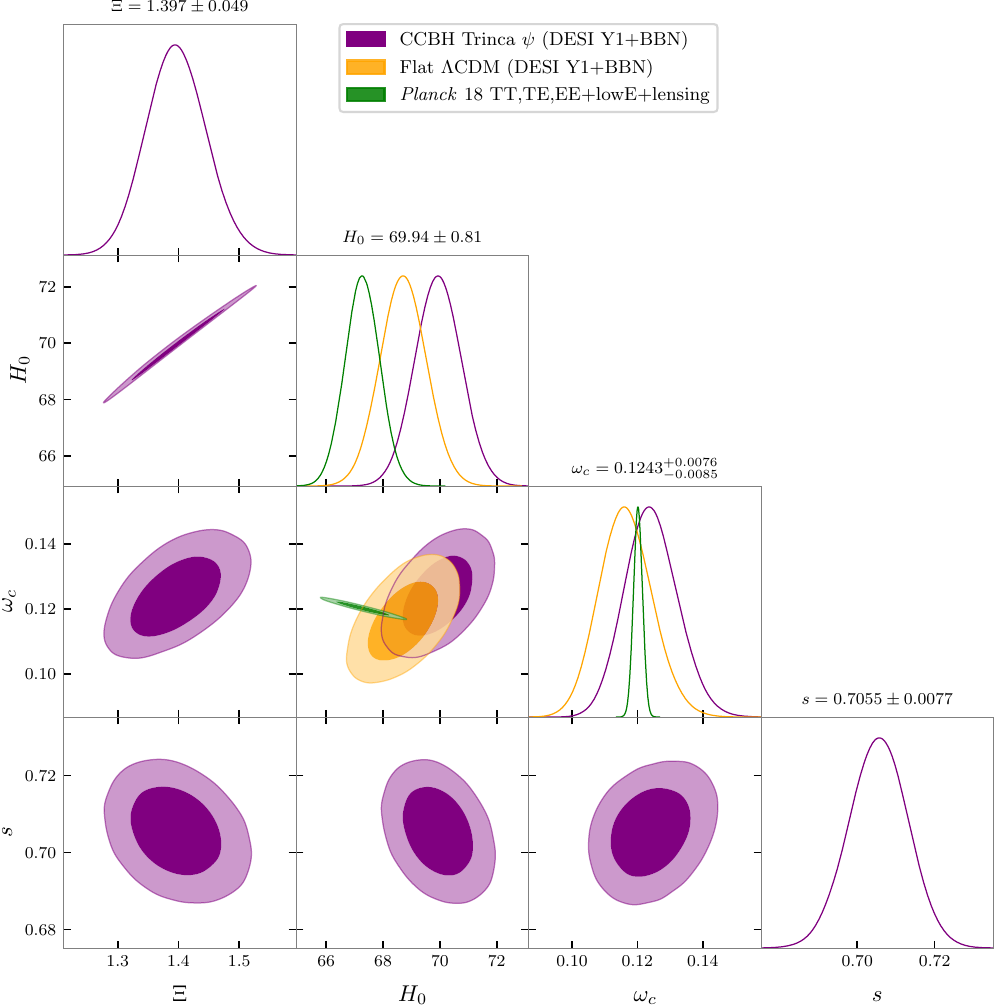}
  \caption{\label{fig:corner} Posterior distributions for DESI BAO+BBN fits to DE sourced by cosmologically coupled BHs (purple), and $\Lambda$CDM (orange).
    \emph{Planck} posteriors are shown (green) for comparison.
    Both fits draw the comoving baryon density inferred from early-universe measurements $\omega_b^\mathrm{proj}$ from a Gaussian BBN prior, which dominates the posterior, and has been omitted for clarity.
    Both fits draw the comoving cold dark matter density $\omega_c$ from the same uninformative prior.
    For CCBH, we draw $\Xi$: the amount of baryonic matter becoming BHs, per unit stellar material; while for $\Lambda$CDM, we draw $H_0$.
    For CCBH, the baryon survival fraction $s$ is also displayed.
    The posterior for $s$ implies that $\sim 30\%$ of baryons are genuinely lost, having been converted into DE inside BHs.
    This present-day baryon abundance is consistent with all contemporary baryon censuses.
  }
\end{figure}%
Our goal is to determine the DESI BAO best-fit background-only CCBH and $\Lambda$CDM cosmologies.
By fitting $\Lambda$CDM, we validate our independent pipeline against that of DESI.
For CCBH, we fit for the collapse fraction $\Xi$, the free parameter that replaces $H_0$ used by DESI when fitting background-only $\Lambda$CDM with a BBN prior.
For our primary analysis, we adopt a redshift $z > 4$ SFRD that includes contributions from intrinsically faint objects \cite[][Fig.~3]{Trinca2024}, while accurately reproducing $M_{UV} < -17$ observations from JWST \cite[e.g.][Fig.~8]{AdamsSFRD2024} (although \cite[c.f.][]{DSilva2023_AGNContamination, DSilva2023_GAMADEVILS}).
For $z \leqslant 4$, we adopt the standard Madau \& Dickinson \cite[][]{MadauDickinson2014, madaufragos2017} SFRD, enforcing continuity by scaling the normalization to match at $z=4$.
The required factor $\sim 2$ is consistent with radio astrometric measures of the SFRD from synchrotron emission in protostellar disks \cite{hopkinsbeacom2006}.
These measurements are free from dust-obscuration systematics that impact all IR, UV, and optical inference of the SFRD and likely better determine the normalization.
This combination of $z > 4$ and $z < 4$ behavior we will designate as ``Trinca $\psi$.''
We additionally compute with the standard Madau \& Dickinson SFRD, designated as ``Madau $\psi$,'' to provide a fiducial visual reference in \cref{fig:sfrd}.%
\footnote{Empirical determinations of the SFRD necessarily assume a particular cosmology to convert observed luminosities into stellar production rates, and to convert object counts into comoving densities.
Because the expansion history does not begin to deviate from $\Lambda$CDM until $z\sim 1$ when star-formation is winding down, and the corrections are small relative to the original data uncertainties, we have neglected this correction.}

Our code does not track recombination thermodynamics, so we adopt \cite[][Eqn.~(16)]{aubourg2015}
\begin{align}
  r_\mathrm{d} = 55.154 \frac{\exp\left[-72.3\left(\omega_\nu + 0.0006\right)^2\right]}{\left(\omega_c + \omega_b^\mathrm{proj}\right)^{0.25351}\left(\omega_b^{\mathrm{proj}}\right)^{0.12807}}~\mathrm{Mpc}\,,
\end{align}
to compute the baryon drag scale $r_\mathrm{d}$ to better than $0.06\%$.%
\footnote{This model for $r_\mathrm{d}$ assumes $N_\mathrm{eff} = 3.046$ and delivers accuracy to $0.021\%$.
This $N_\mathrm{eff}$ is larger than our adopted $N_\mathrm{eff} = 3.044$ by $0.06\%$, so we report the larger uncertainty.}
Next, define:
\begin{align}
  D_H(z) := \frac{c}{H(z)} \qquad\qquad D_V(z) := \left[z D_\mathrm{M}(z)^2D_H(z)\right]^{1/3}.
\end{align}
DESI can extract a BAO scale, relative to $r_\mathrm{d}$, in three ways: 1) fixed redshift and variable sky position via the transverse comoving distance $D_\mathrm{M}/r_\mathrm{d}$; 2) fixed sky position and variable redshift via the Hubble distance $D_H/r_\mathrm{d}$; and 3) the ``angle averaged'' mixture of these two signals $D_V/r_\mathrm{d}$ when statistics are not sufficient to report $D_\mathrm{M}/r_\mathrm{d}$ and $D_H/r_\mathrm{d}$ separately.
The DESI-1YR BAO data include five distinct measurements each of $D_\mathrm{M}/r_\mathrm{d}$, $D_H/r_\mathrm{d}$, and two of $D_V/r_\mathrm{d}$ (see \cref{fig:datafits}).
The best-fit central values $\mathbf{\mu}$ and covariance matrix for these 12 data points $\Sigma$ is provided by the DESI collaboration.%
\footnote{This data can be found at \url{https://github.com/CobayaSampler/bao_data/tree/master}, and is significantly more precise than the data as listed in DESI Table 1.}
We solve \cref{eqn:ode_system} with \texttt{scipy.odeint} to determine model-predicted values $\mathbf{X}$ for the these 12 data points at their respective redshifts of measurement $z_\mathrm{eff}$.
Because the variables are Gaussian distributed,
\begin{align}
  \log \mathcal{L}_\mathrm{DESI}(\mathbf{X}) := -\frac{1}{2} (\mathbf{X} - \mathbf{\mu})^\top \Sigma^{-1} (\mathbf{X} - \mathbf{\mu}),
\end{align}
where $\top$ denotes transpose and $\Sigma^{-1}$ denotes the inverse covariance matrix.
We additionally enforce a hard cutoff in late-time baryon density based on confirmed baryon census in the local universe \cite{Driver2021},
\begin{align}
  \log \mathcal{L}_b\left(s\right) :=
  \begin{cases}
    0 & s > 0.25 \\
    -\infty & s \leqslant 0.25
    \end{cases},
\end{align}
where $s$ is the baryon survival fraction,
\begin{align}
  s := \frac{\rho_b(1)}{\rho_b(a_i)a_i^3}.
\end{align}
The combined $\log$ likelihood becomes
\begin{align}
  \log \mathcal{L} := \log \mathcal{L}_\mathrm{DESI} + \log \mathcal{L}_b. \label{eqn:logL}
\end{align}

For consistency, we assert the same priors for CCBHs and $\Lambda$CDM: the DESI-adopted BBN prior on $\omega_b^\mathrm{proj}$ and a uniform prior in $\omega_c$.
  These are described in Table~\ref{tbl:fit_parameters}.
  Although DESI fits using $\Omega_m$, our $\Lambda$CDM posterior distributions agree with DESI to better than $0.5\%$ precision.%
  \footnote{We verified our pipeline to $<0.02\%$ precision against \texttt{astropy}, which is $25\times$ smaller than $0.5\%$.
  To investigate this discrepancy, we constructed a secondary pipeline by modifying a version of \texttt{CLASS} \cite{BlasLesgourges2011}.
  A \texttt{MontePython} \cite{Audren:2012wb, Brinckmann:2018cvx} fit using an independently developed DESI Y1 BAO likelihood recovered the same values as our primary pipeline to $< 0.02\%$ precision.
  We conclude our inference to be robust.}
In $\Lambda$CDM, the (fixed) DE density is determined by flatness and choice of $H_0$.
For CCBHs, the DE density evolves away from zero according to \cref{eqn:ode_system}.
Thus, at our background-order study, CCBHs (given some adopted $\psi$) and $\Lambda$CDM have the same number of free parameters, but CCBHs give time-evolution with two fewer parameters than $w_0w_a$ models.

We use the dynamic nested sampling framework \texttt{dynesty} \cite{2020MNRAS.493.3132S, 2004AIPC..735..395S, 10.1214/06-BA127, 2019S&C....29..891H, 2009MNRAS.398.1601F}%
\footnote{The version we used can be found at \url{doi.org/10.5281/zenodo.3348367}}%
,  to measure the Bayesian evidence for \cref{eqn:logL} and estimate posterior distributions for $\omega_b^\mathrm{proj}$, $\omega_c$, $H_0$ and $\Xi$ for each cosmology. 
With these distributions, we compute derived distributions for $H_0$, $\Omega_m$, and the survival fraction $s$.
  The survival fraction for $\Lambda$CDM is expected to be unity to integrator precision, and serves as a further internal consistency check on our primary pipeline.
  We find $s_{\Lambda\mathrm{CDM}} = 1.00015$, in agreement with expectations.

\section{Results}
\label{sec:results}
\cref{fig:sfrd} shows the evolution of DE density $\rho_\mathrm{DE}$ for CCBH cosmologies, adopting the two SFRD models described in \cref{sec:methods}.
Overlaid in color are the $\pm 1\sigma$ regions implied by DESI $w_0w_a$ parameters for each of the three Planck+DESI+SN data sets.
Note that the SNe regions have the ``hump'' shape and the CCBH curves have the asymptoting ``plateau'' shape, as expected from monotonically decreasing star-formation for $z < 2$.
\cref{tbl:fit_parameters} summarizes all best-fit parameter values, $\chi^2 := -2 \log \mathcal{L}$ for the best-fits, and posterior means with $68\%$ confidence regions.
For CCBH Trinca $\psi$, the best-fit parameters all lie within $\pm 1\sigma$ of their posterior means.
For flat $\Lambda$CDM, the best-fit $H_0$ and $\omega_b$ pull away from the posterior mean at $<-1\sigma$ and $<-2\sigma$, respectively.
Because the posterior is the prior-weighted likelihood by Bayes' Theorem, the discrepancy in $\omega_b$ indicates that the data prefer a lower comoving baryon density than Flat $\Lambda$CDM can accommodate.

\cref{fig:datafits} shows our determined distance measures $D_\mathrm{M}/r_d$, $D_H/r_d$, and $D_V/r_d$, with Flat $\Lambda$CDM (\emph{Planck}) distance measures subtracted off.
It is not yet possible to distinguish late-time Flat $\Lambda$CDM (DESI 1YR+BBN) from CCBH Trinca $\psi$ (DESI 1YR+BBN) at background-order.
DESI reports that their (late-time) $\Lambda$CDM is, however, distinct from \emph{Planck} (early-time) $\Lambda$CDM at $1.9\sigma$.
If this discrepancy should increase, then the $\Lambda$CDM model will be unable to reconcile early and late universe cosmological probes.
However, CCBH can provide the missing late-time physics to reconcile these two datasets.
Improved statistics from DESI data, and a comprehensive first-order study of CCBH with the \emph{Planck} likelihoods, will help clarify whether or not this is the case.

\cref{fig:corner} summarizes the posterior distributions from our primary CCBH Trinca $\psi$ and Flat $\Lambda$CDM analyses.
For comparison, we also include \emph{Planck} TT,TE,EE+lowE+lensing posteriors.
We have omitted $\omega_b^\mathrm{proj}$ because the posterior distribution for $\omega_b^\mathrm{proj}$ tracks the prior (see \cref{tbl:fit_parameters}).
The present-day expansion rate $H_0 = (69.94\pm 0.81)\,\kmsMpc$ is notably higher than both Flat $\Lambda$CDM (DESI 1YR+BBN) and the \emph{Planck} value, because the consumption of baryons allows earlier transition to DE domination.
With respect to the latter, CCBH reduces gaussian tension in $H_0$ from $5.6\sigma$ to $2.7\sigma$ when compared to the SH0ES measurement \cite{Murakami23}.
We highlight that this value agrees to $0.5\%$ with $H_0 = (69.59 \pm 1.58)\,\kmsMpc$ reported by the Chicago-Carnegie Hubble Program using Cepheids, the Tip of the Red Giant Branch (TRGB), and the J-Region Asymptotic Giant Branch stellar distance-ladder calibrations \cite{2024arXiv240806153F}.
In the CCBH Trinca $\psi$ cosmology, the recovered value for $\omega_c$ is higher than \emph{Planck}, allowing this CCBH cosmology to accommodate more background matter density contribution from massive neutrinos.
In contrast, the Flat $\Lambda$CDM cosmology requires a lower $\omega_c$ than \emph{Planck} in order to achieve the smaller $\Omega_m$ preferred by DESI BAO.

\section{Discussion}
\label{sec:discussion}
We have reproduced the BAO data measured by DESI with a model that does not require adjustments to early-universe physics, and that leverages a well-motivated astrophysical source for DE: stellar-collapse CCBHs.
This is the only known model that explains the coincidence problem of why DE has become relevant now: because it didn't exist until stars formed.
This also explains why the DE density is of the same order of magnitude as the matter density: DE is sourced by matter.
Physically, $\Xi = 1.39$ implies that, on average, $1.39M_\odot$ of baryons convert into BH DE for every $1M_\odot$ of stellar material produced.
This value of $\Xi$ is recovered when $s = 70\%$, implying that approximately $30\%$ of baryons are missing at late times.
In the CCBH model, this means that those baryons are truly lost, being converted into DE.

In fact, there has been a well-known ``missing baryons problem'' for over a decade, with measurements indicating a shortfall of about $30\%$ \cite{Shull2012}.
Our estimated survival fraction is consistent with all contemporary baryon censuses \cite[][Fig.~1]{Driver2021}. 
This consumption of baryons may help explain other unexpected aspects of the DESI-1YR analysis.
DESI-1YR results find a most probable summed neutrino mass $\sum m_\nu$ equal to zero \cite[Fig.~11, Left Panel][]{desicollaboration2024desi}(see also \cite{Wang:2024hen}), contradicting the computed $\sum m_\nu > 0.059~\mathrm{eV}$ based on $\Delta m_{ij}^2 $ constraints from neutrino oscillation measurements \cite[e.g.][]{dayabay, 2021Univ....7..459G}.%
\footnote{This standard constraint follows from algebra
\begin{align}
  \sqrt{\Delta m_{21}^2} + \sqrt{\Delta m_{31}^2} = (m_2 - m_1) + (m_3 - m_1) < m_2 + m_3 < \sum m_\nu,
\end{align}
under the assumption that $m_\nu \geqslant 0~\forall \nu$ and $m_i$ are monotonically increasing in $i$ (normal hierarchy).}
The summed neutrino mass approaching zero in DESI-1YR can be interpreted as a preference for lower matter density at late times, relative to that inferred from the CMB.%
\footnote{It has also been proposed that a preferred zero summed neutrino mass could arise from new physics in the neutrino sector \cite{Craig:2024tky}}
We suspect that the baryon consumption required in the BH DE scenario will allow $\sum m_\nu$ to increase towards physically reasonable values.
At background order, $\Omega_bh^2$ and $\sum m_\nu$ are not entirely degenerate in a CCBH cosmology, as the required DE density constrains the amount of baryon consumption.
We note that $\sum m_\nu$ is also sensitive to the growth function at first-order in cosmological perturbations, which can be altered by novel CCBH dynamics \cite[e.g.][\S4]{CrokerRunburg20}. 
We further expect that the decreased baryon density along the line of sight will slightly increase $z_\mathrm{reio}$, because $\tau_\mathrm{reio}$ is predominantly determined by the low $\ell$ multipoles of the CMB EE polarization power-spectrum.
A comprehensive investigation of these, and other first-order observables, is the subject of future work.

To place the measured value of $\Xi$ into context, we assume a Chabrier \cite{ChabrierIMF2003} distribution of stellar masses at birth $\d N/\d m \propto m^{-2.3}$.
Then the fraction of stellar mass in stars large enough to form BHs ($m \gtrsim 20M_\odot$) is $f = 0.4$.
A $\Xi = 1.4$ implies approximately $3.45\times$ additional baryonic consumption, which is reasonable given astrophysical uncertainties.
For example, the stellar Initial Mass Function (IMF) at $z \gtrsim 1$ is an active area of investigation \citep[e.g.][]{SneppenSteinhardt2022, SteinhardtSneppen2022}. 
If the IMF at $z\gtrsim1$ were `top-heavy,' as suggested by some studies \citep{baugh05,lacey16,iani23,yung24}, the required factor decreases.
For example, an IMF $\propto m^{-2.1}$ \citep[e.g.][]{2023Natur.613..460L} gives $f = 0.74$, reducing the factor to $1.86\times$.
Post-collapse accretion could account for part of $\Xi$, though estimating its extent is challenging. 
Accretion onto isolated BHs is believed to be inefficient.
Because $\Xi$ describes average behavior, a scenario where the high-end of the BH mass function accretes by a factor of $\sim 10^2\text{\textendash} 10^3\times$, while the low end evolves passively, is conceivable.
Gravitational processes can also produce mass.
This physics is independent of $\Xi$ because \cref{eqn:background_cons} has no knowledge of local gravitational processes.
The Press and Teukolsky process \cite{Misner1972, PressTeukolsky1972} can convert BH spin into mass, provided the radiated energy is absorbed by further infalling material during core collapse.
Local strong gravitational effects would effectively amplify the term on the right-hand side of \cref{eqn:background_cons}, without altering the depletion described by \cref{eqn:rho_b_defn}.
We note that such effects may involve mediating fields, such as scalar fields \cite{SilvaSakstein2018, Herdeiro2021}.

We have focused on a ``cosmic noon'' scenario, but production of DE at the peak of star-formation complements production near ``cosmic dawn'' $z \sim 20$, as has been previously studied \cite{2018PhDT.......194C, CrokerWeiner19, CrokerRunburg20, Farrah23b}.
The impact of Pop III \cite[e.g.][]{MaiolinoPopIIIEvidence2024} and direct-collapse BHs \cite[e.g.][]{Nabizadeh2024} is to establish an effective DE ``floor.''
Production of this DE floor does not significantly deplete the baryon density, because the DE produced per unit baryon density scales as $(1+z)^3$.
For example, in \emph{Planck} 2018 cosmologies, $\omega_\Lambda \sim 5\omega_c/2$.
Assume a cosmic dawn burst of production, either direct collapse or Pop III stellar collapse, at $z_\mathrm{III}$.
Then
\begin{align}
  \frac{5\omega_c}{2} = \omega_b\left(1-s_\mathrm{III}\right)\left(1+z_\mathrm{III}\right)^3.
\end{align}
Adopting $z_\mathrm{III} = 15$ then gives a baryon survival of $s_\mathrm{III} = 99.67\%$, so consumption of $0.4\%$ of baryons is sufficient to produce \emph{all} required DE.
Any DE floor established at high $z$ would further decrease $\Xi$.

In summary, DESI’s first year of observations provide evidence that DE evolves over time using a $w_0 w_a$ parameterization.
If DE is produced from the conversion of baryonic matter into cosmologically coupled BHs, the DE density will increase as massive stars form and collapse into BHs, rising from zero toward a plateau as star-formation quenches at $z \lesssim 2$. 
We have tested this hypothesis by fitting a suitable Friedmann cosmology to the DESI measured Baryon Acoustic Oscillation signals between $0 < z < 2.5$.
Our calculated evolving DE density agrees with the DESI $w_0w_a$ models at $< 1\sigma$, except at redshifts $z \lesssim 0.2$, where the $w_0w_a$ parameterization becomes inadequate for our phenomenology.
Cosmologically coupled BHs produce $H_0 = (69.94 \pm 0.81) \kmsMpc$, reducing tension with SH0ES and in excellent agreement with $H_0 = (69.59 \pm 1.58)\,\kmsMpc$ reported recently by Chicago-Carnegie Hubble Program using Cepheids, Tip of the Red Giant Branch (TRGB), and J-Region Asymptotic Giant Branch stellar distance-ladder calibrations.
They provide a physical explanation for this larger central value: the Hubble rate today is higher than that inferred from the Cosmic Microwave Background because the conversion of baryons into DE during star-formation causes an earlier transition from matter to DE dominance.
In addition to reducing tensions between early and late-time observations, the scenario we have investigated provides an astrophysically motivated explanation for a known $\sim 30\%$ deficit in the present-day baryon density, relative to BBN expectations.
Furthermore, the conversion of baryons into DE decreases the present-day matter density of Friedmann models, allowing the sum of neutrino masses to increase away from zero.
No other current DE model is both astrophysically motivated and capable of reconciling early and late universe measurements of the expansion rate.

\begin{acknowledgments}
  The authors thank the anonymous referee for very useful comments that resulted in substantial improvement of the manuscript.
  We further thank Arnaud de Mattia and Seshadri Nadathur (DESI) for discussion concerning regeneration of DESI results, Thomas Tram (Norway) for guidance with massive relics in \texttt{CLASS}, Mark Dickinson (NOIRLab) for guidance with cosmological corrections to SFRDs, and Christopher Cain (ASU) for input on reionization.
  KC~thanks the Sakai Group at Yamaguchi University for their hospitality during the preparation of this work, Ryo Saito (Yamaguchi) for comments on $w_0w_a$ priors, Joel Weiner (Hawai`i) for discussion, and Tom Browder (Hawai`i) for comments on neutrinos.
  GT~acknowledges support through DoE Award DE-SC009193.
  RAW~acknowledges support from NASA JWST Interdisciplinary Scientist grants NAG5-12460, NNX14AN10G and 80NSSC18K0200 from GSFC. The work of NF~is supported by DOE grant DOE-SC0010008. 
  NF~thanks the Aspen Center for Physics, which is supported by the National Science Foundation grant PHY-2210452, and the Sloan Foundation for its partial support.
\end{acknowledgments}

\begin{appendix}

\end{appendix}
\bibliography{main}{}
\bibliographystyle{JHEP}

\end{document}